\documentclass[a4paper,10pt]{article}
\pdfoutput=1
\usepackage[utf8]{inputenc}
\usepackage[a4paper, total={7in, 9in}]{geometry}

\usepackage[utf8]{inputenc}
\usepackage[T1]{fontenc}
\usepackage{subfiles}
\usepackage[english]{babel} 
\usepackage{framed}
\usepackage{amsmath}
\usepackage{amssymb}
\usepackage{listings}
\usepackage{graphicx}
\usepackage{afterpage}
\usepackage{mathtools}
\usepackage{xspace}
\usepackage{siunitx}
\usepackage{caption}
\usepackage{subcaption}
\usepackage{hyperref}
\usepackage{pgfplots}
\usepackage{pgfplotstable}
\usepackage{booktabs}
\usepackage{natbib}

\usepackage[labelsep=period]{caption}
\usepackage{titlesec}

\titleformat*{\subsection}{\large\itshape}

\newcommand{\Hm}{\ensuremath{\vec{H}^\star}} 
\newcommand{\Bm}{\ensuremath{\vec{B}^\star}}
\newcommand{\qd}{\mathrm{qd}}

\usepackage{glossaries}  
\newacronym{cad}{CAD}{computer-aided design}
\newacronym[longplural=degrees of freedom]{dof}{DoF}{degree of freedom}
\newacronym{fe}{FE}{finite element}
\newacronym{fem}{FEM}{finite element method}
\newacronym{fit}{FIT}{finite integration technique}
\newacronym{nurbs}{NURBS}{non-uniform rational B-splines}
\newacronym{pec}{PEC}{perfect electric conductor}
\newacronym{pmc}{PMC}{perfect magnetic conductor}
\newacronym{rf}{RF}{radio-frequency}
\newacronym{te}{TE}{transverse electric}
\newacronym{tm}{TM}{transverse magnetic}
\newacronym{tem}{TEM}{transverse electric and magnetic}
\newacronym[longplural=boundary conditions]{bc}{BC}{boundary condition}

\newacronym{uq}{UQ}{uncertainty quantification}
\newacronym{pml}{PML}{perfectly matched layers}
\newacronym{fdtd}{FDTD}{finite-difference time-domain}
\newacronym{dgtd}{DGTD}{discontinuous Galerkin time-domain}
\newacronym{mim}{MIM}{metal-insulator-metal}
\newacronym{rhs}{RHS}{right-hand side}
\newacronym{sqp}{SQP}{sequential quadratic programming}
\newacronym{lar}{LAR}{least angle regression}
\newacronym{td}{TD}{total degree}
\newacronym{tp}{TP}{tensor product}
\newacronym{hc}{HC}{hyperbolic cross}
\newacronym{lhs}{LHS}{latin hypercube sampling}
\newacronym{ls}{LS}{least squares}
\newacronym{qmc}{QMC}{quasi Monte Carlo}
\newacronym{mlmc}{MLMC}{multilevel Monte Carlo}
\newacronym{hf}{HF}{high-frequency}
\newacronym{megpc}{MEGPC}{multi-element generalized polynomial chaos}
\newacronym{gp}{GP}{Gaussian process}
\newacronym{ed}{ED}{experimental design}
\newacronym{rms}{RMS}{root mean square}
\newacronym{cv}{CV}{cross-validation}
\newacronym{es}{ES}{educated sampling}
\newacronym{iga}{IGA}{isogeometric analysis}
\newacronym{bem}{BEM}{boundary element method}
\newacronym{ibvp}{IBVP}{initial/boundary value problem}
\newacronym{em}{EM}{electromagnetic}
\newacronym{emf}{EMF}{electromagnetic field}
\newacronym{gpc}{gPC}{generalized polynomial chaos}
\newacronym{mc}{MC}{Monte Carlo}
\newacronym{pde}{PDE}{partial differential equation}
\newacronym{pdf}{PDF}{probability density function}
\newacronym[longplural={quantities of interest}]{qoi}{QoI}{quantity of interest}
\newacronym{rv}{RV}{random variable}
\newacronym{ae}{a.e.}{almost every}

\DeclareMathOperator*{\argmin}{\arg\!\min}

\newcommand{\drawTD}[2]
{
        \foreach \x in {0,1,2,...,#1} {
            \foreach \y in {0,1,2,...,#1} {
				\ifnum \numexpr\x+\y < \numexpr#1+1 
               		\node at (\x,\y) [circle, fill=#2, minimum size=5mm] {};
             	\fi 
            }
        }
}

\def\addlegendimage{\csname pgfplots@addlegendimage\endcsname}

\newcommand{\drawCoordinateSystem}[4] % xmin,xmax,ymax,circle
{
	\draw[thick,->] (#1+#4,0) -- (#2,0) node[anchor=south] {$z$};
	\draw[thick,->] (#1,#4) -- (#1,#3) node[anchor=north west] {$y$};
	\draw[thick] (#1,0) circle (#4);
	\draw[thick] (#1-0.707*#4,-0.707*#4) -- (#1+0.707*#4,0.707*#4){};
	\draw[thick] (#1-0.707*#4,0.707*#4) -- (#1+0.707*#4,-0.707*#4)node[anchor=south west] {$x$};
}

\pgfmathdeclarefunction{gauss}{2}{%
  \pgfmathparse{1/(#2*sqrt(2*pi))*exp(-((x-#1)^2)/(2*#2^2))}%
}

\tikzset{%
	remember picture with id/.style={%
		remember picture,
		overlay,
		save picture id=#1,
	},
	save picture id/.code={%
		\edef\pgf@temp{#1}%
		\immediate\write\pgfutil@auxout{%
			\noexpand\savepointas{\pgf@temp}{\pgfpictureid}}%
	},
	if picture id/.code args={#1#2#3}{%
		\@ifundefined{save@pt@#1}{%
			\pgfkeysalso{#3}%
		}{
			\pgfkeysalso{#2}%
		}
	}
}

\def\savepointas#1#2{%
	\expandafter\gdef\csname save@pt@#1\endcsname{#2}%
}

\def\tmk@labeldef#1,#2\@nil{%
	\def\tmk@label{#1}%
	\def\tmk@def{#2}%
}

\tikzdeclarecoordinatesystem{pic}{%
	\pgfutil@in@,{#1}%
	\ifpgfutil@in@%
	\tmk@labeldef#1\@nil
	\else
	\tmk@labeldef#1,(0pt,0pt)\@nil
	\fi
	\@ifundefined{save@pt@\tmk@label}{%
		\tikz@scan@one@point\pgfutil@firstofone\tmk@def
	}{%
		\pgfsys@getposition{\csname save@pt@\tmk@label\endcsname}\save@orig@pic%
		\pgfsys@getposition{\pgfpictureid}\save@this@pic%
		\pgf@process{\pgfpointorigin\save@this@pic}%
		\pgf@xa=\pgf@x
		\pgf@ya=\pgf@y
		\pgf@process{\pgfpointorigin\save@orig@pic}%
		\advance\pgf@x by -\pgf@xa
		\advance\pgf@y by -\pgf@ya
	}%
}

\makeatother
\definecolor{TUDa-0d}{cmyk/RGB/HTML}{0,0,0,.8/83,83,83/535353}
\definecolor{TUDa-0c}{cmyk/RGB/HTML}{0,0,0,.6/137,137,137/898989}
\definecolor{TUDa-0b}{cmyk/RGB/HTML}{0,0,0,.4/181,181,181/B5B5B5}
\definecolor{TUDa-0a}{cmyk/RGB/HTML}{0,0,0,.2/220,220,220/DCDCDC}

\definecolor{TUDa-1a}{cmyk/RGB/HTML}{.7,.4,0,0/93,133,195/5D85C3}
\definecolor{TUDa-2a}{cmyk/RGB/HTML}{0.8,.2,0,0/0,156,218/009CDA}
\definecolor{TUDa-3a}{cmyk/RGB/HTML}{0.7,0,.5,0/80,182,149/50B695}
\definecolor{TUDa-4a}{cmyk/RGB/HTML}{.4,0,.8,0/175,204,80/AFCC50}
\definecolor{TUDa-5a}{cmyk/RGB/HTML}{.2,0,.8,0/221,223,72/DDDF48}
\definecolor{TUDa-6a}{cmyk/RGB/HTML}{0,.1,.7,0/255,224,92/FFE05C}
\definecolor{TUDa-7a}{cmyk/RGB/HTML}{0,.3,.8,0/248,186,60/F8BA3C}
\definecolor{TUDa-8a}{cmyk/RGB/HTML}{0,.6,.8,0 /238,122,52/EE7A34}
\definecolor{TUDa-9a}{cmyk/RGB/HTML}{0,.8,.7,0/233,80,62/E9503E}
\definecolor{TUDa-10a}{cmyk/RGB/HTML}{.2,.9,0,0/201,48,142/C9308E}
\definecolor{TUDa-11a}{cmyk/RGB/HTML}{.6,.8,0,0/128,69,151/804597}
\definecolor{TUDa-1b}{cmyk/RGB/HTML}{1,.6,0,0/0,90,169/005AA9}
\definecolor{TUDa-2b}{cmyk/RGB/HTML}{1,.3,0,0/0,131,204/0083CC}
\definecolor{TUDa-3b}{cmyk/RGB/HTML}{1,0,.6,0/0,157,129/009D81}
\definecolor{TUDa-4b}{cmyk/RGB/HTML}{.5,0,1,0/153,192,0/99C000}
\definecolor{TUDa-5b}{cmyk/RGB/HTML}{.3,0,1,0/201,212,0/C9D400}
\definecolor{TUDa-6b}{cmyk/RGB/HTML}{0,.2,1,0/253,202,0/FDCA00}
\definecolor{TUDa-7b}{cmyk/RGB/HTML}{0,.4,1,0/245,163,0/F5A300}
\definecolor{TUDa-8b}{cmyk/RGB/HTML}{0,.7,1,0/236,101,0/EC6500}
\definecolor{TUDa-9b}{cmyk/RGB/HTML}{0,1,.9,0/230,0,26/E6001A}
\definecolor{TUDa-10b}{cmyk/RGB/HTML}{.4,1,0,0/166,0,132/A60084}
\definecolor{TUDa-11b}{cmyk/RGB/HTML}{.7,1,0,0/114,16,133/721085}
\definecolor{TUDa-1c}{cmyk/RGB/HTML}{1,.7,.2,0/0,78,138/004E8A}
\definecolor{TUDa-2c}{cmyk/RGB/HTML}{1,.5,.2,0/0,104,157/00689D}
\definecolor{TUDa-3c}{cmyk/RGB/HTML}{1,.2,.6,0/0,136,119/008877}
\definecolor{TUDa-4c}{cmyk/RGB/HTML}{.6,.1,1,0/127,171,22/7FAB16}
\definecolor{TUDa-5c}{cmyk/RGB/HTML}{.4,.1,1,0/177,189,0/B1BD00}
\definecolor{TUDa-6c}{cmyk/RGB/HTML}{.2,.3,1,0/215,172,0/D7AC00}
\definecolor{TUDa-7c}{cmyk/RGB/HTML}{.2,.5,1,0/210,135,0/D28700}
\definecolor{TUDa-8c}{cmyk/RGB/HTML}{.2,.8,1,0/204,76,3/CC4C03}
\definecolor{TUDa-9c}{cmyk/RGB/HTML}{.3,1,.9,0/185,15,34/B90F22}
\definecolor{TUDa-10c}{cmyk/RGB/HTML}{.5,1,.3,0/149,17,105/951169}
\definecolor{TUDa-11c}{cmyk/RGB/HTML}{.8,1,.2,0/97,28,115/611C73}
\definecolor{TUDa-1d}{cmyk/RGB/HTML}{1,.9,.3,0/36,53,114/243572}
\definecolor{TUDa-2d}{cmyk/RGB/HTML}{1,.7,.4,0/0,78,115/004E73}
\definecolor{TUDa-3d}{cmyk/RGB/HTML}{1,.4,.7,0/0,113,94/00715E}
\definecolor{TUDa-4d}{cmyk/RGB/HTML}{.7,.3,1,0/106,139,55/6A8B22}
\definecolor{TUDa-5d}{cmyk/RGB/HTML}{.5,.2,1,0/153,166,4/99A604}
\definecolor{TUDa-6d}{cmyk/RGB/HTML}{.4,.4,1,0/174,142,0/AE8E00}
\definecolor{TUDa-7d}{cmyk/RGB/HTML}{.3,.6,1,0/190,111,0/BE6F00}
\definecolor{TUDa-8d}{cmyk/RGB/HTML}{.4,.8,1,0/169,73,19/A94913}
\definecolor{TUDa-9d}{cmyk/RGB/HTML}{.5,1,.9,0/156,28,38/961C26}
\definecolor{TUDa-10d}{cmyk/RGB/HTML}{.7,1,.5,0/115,32,84/732054}
\definecolor{TUDa-11d}{cmyk/RGB/HTML}{.9,1,.3,0/76,34,106/4C226A}

\newcommand{\rev}[1]{\textcolor{black}{#1}}

\title{\textbf{Three-Dimensional Data-Driven \rev{Magnetostatic} Field \rev{Computation} using Real-World Measurement Data}}

\author{Armin Galetzka$^1$, Dimitrios Loukrezis$^{1,2}$, Herbert De Gersem$^{1,2}$}

\date{
	\small{$^1$\emph{Technische Universit\"at Darmstadt, Institute for Accelerator Science and Electromagnetic Fields (TEMF)} \\ \emph{Schlossgartenstrasse 8, 64289 Darmstadt, Germany}} \\ 
	\small{$^2$\emph{Technische Universit\"at Darmstadt, Centre for Computational Engineering} \\ \emph{Dolivostrasse 15, 64293 Darmstadt, Germany}} \\ 
}

\begin{document}
\maketitle
\begin{abstract}
	\noindent \textbf{Purpose} - The purpose of this paper is to present the applicability of data-driven solvers to computationally demanding three-dimensional problems and their practical usability when utilizing real-world measurement data.\\
	\textbf{Design/methodology/approach} - Instead of using a hard-coded phenomenological material model within the solver, the data-driven computing approach reformulates the boundary value problem such that the field solution is directly computed on raw measurement data. The data-driven formulation results in a double minimization problem based on Lagrange multipliers, where the sought solution must conform to Maxwell's equations while at the same time being as close as possible to the available measurement data. The data-driven solver is applied to a three-dimensional model of a \rev{DC-current electromagnet}. \\
	\textbf{Findings} - Numerical results for data sets of increasing cardinality verify that the data-driven solver recovers the conventional solution. Additionally, the practical usability of the solver is shown by utilizing real-world measurement data. This work concludes that the data-driven magnetostatic finite element solver is applicable to computationally demanding three dimensional problems, as well as in cases where a prescribed material model is not available. \\
	\textbf{Originality} - While the mathematical derivation of the data-driven problem is well presented in the referenced papers, the application to computationally demanding real-world problems, including real measurement data and its rigorous discussion is missing. The presented work closes this gap and shows the applicability of data-driven solvers to challenging, real-world test cases. 
	\textbf{Paper type }Research Paper
\end{abstract}

\section{Introduction}
A large class of electromagnetic field problems are described by \glspl{ibvp}, where Maxwell's equations relate fields to their sources and constitutive laws describe material behavior. Constitutive laws are commonly built upon two ingredients. First, a sufficient amount of measurement data must be provided. In the case of magnetostatics, the data consist of measured $(BH)$-pairs. Second, a model is selected, i.e. a closed-form relation with yet unknown coefficients that is thought to represent the material behavior. The material model is then obtained through data-fitting techniques, e.g. machine learning regression \cite{grech2020} or physics-preserving methods \cite{pechstein2006}.
While Maxwell's laws are accepted as exactly known, data-fitted material models suffer from so-called epistemic uncertainty, in the sense that the chosen model may arbitrarily deviate from the actual material behavior. 
In turn, this uncertainty propagates to the numerical solutions of solvers with fixed material models.

Recent works \rev{in the field of computational mechanics} circumvent the demand for a closed-form material model altogether and instead solve the \gls{ibvp} directly on the measurement data \cite{kirchdoerfer2016data, kirchdoerfer2017data}. 
Therefore, this so-called data-driven computing framework avoids material modeling errors and epistemic uncertainties as well. 
\rev{Furthermore, the data-driven solver is also applicable in cases where no prescribed model is available or an adequate choice of a model is not possible. Such cases appear for example if a new material is measured and established models fail to cover the entire regime of the material's behaviour properly. Constructed models may also fail entirely, in the sense that numerical solvers do not converge. For example, in the case where only noisy data is available, a standard interpolation model is impractical. Contrarily, for material models constructed with regression techniques, the convergence of nonlinear \gls{fem} solvers is not necessarily guaranteed and special effort has to be taken into account \cite{pechstein2006}. Examples where conventional material models fail can be found in \cite{ayensa2018new}.}
Due to the fact that the measurement data are directly incorporated into the solver, data-driven computing is particularly suitable in cases of complicated or new materials, where an accurate material law cannot be derived using physical considerations. 
%\rev{Data-driven solvers have been successfully applied on various problems arising in the field of computational mechanics, e.g. fracture mechanics \cite{Carrara2020_fracture}, inelasticity \cite{eggersmann2019model} or for the derviation of material responses \cite{leygue2018data}, to name but a few. Furthermore, effort is spent to reduce the computational costs of the data-driven algorithms and make it a compatible alternative to standard \gls{fem} solvers, see \cite{Kanno2019,Korzeniowski2021}. Yet, the field of electromagnetic field problems, remains mostly underrepresented/unexplored.}

In this work, we build upon existing data-driven electromagnetic field solvers \cite{degersem2020magnetic, galetzka2020datadriven} and extend their application to the case of computationally demanding three-dimensional simulations, as well as to the case of real-world measurement data. 
We consider a highly non-trivial example, namely the model of a \rev{DC-current electromagnet}.
The data-driven formulation for the case of magnetostatics is presented in Section~\ref{sec:dd_framework}. Numerical results on the accuracy and efficiency of the data-driven solver are presented in Section~\ref{sec:numerical_results}. To showcase the solver's applicability in practical test cases, in Section~\ref{sec:numerical_results_real_world} it is utilized in combination with real-world measurement data.
Concluding remarks are available in Section~\ref{sec:conclusion}.
\section{Data-driven framework}
\label{sec:dd_framework}
In the following, we briefly introduce the data-driven solver employed in this work. For a detailed exposition, see \cite{galetzka2020datadriven}. 

We start with the governing equations in magnetostatics, i.e.
\begin{subequations}
	\begin{alignat}{2}
		\mathrm{curl}\,\vec{H} &= \vec{J}, \quad &&\mathrm{in~} \Omega, \\
		\mathrm{div}\,\vec{B} &= 0, \quad &&\mathrm{in~} \Omega, \\
		\vec{B} \cdot \vec{n} &= 0, \quad &&\mathrm{on~} \Gamma,
	\end{alignat}
\label{eq:magnetostatic}%
\end{subequations}%
where $\vec{H}$ is the magnetic field strength, $\vec{B}$ the magnetic flux density, $\vec{J}$ the current density, and $\vec{n}$ the outer unit normal vector.
A homogeneous electric \gls{bc} is assumed on the entire boundary $\Gamma$ of the domain $\Omega$.

In the data-driven framework, we do not rely on a constitutive law between $\vec{H}$ and $\vec{B}$. 
Instead, all information on the material is given through the measurement data set
\begin{equation}
	\widetilde{\mathcal{D}} = \left\{\left(\vec{H}_n^\star,\vec{B}_n^\star\right) \right\}_{n=1}^N.
\label{eq:measurement_data}
\end{equation}
The goal is to minimize the distance between field states $\zeta=(\vec{H},\vec{B})$ that fulfill  \eqref{eq:magnetostatic} and states that conform with the measurement data \eqref{eq:measurement_data}. 
To that end, we collect all Maxwell-conforming states in the set
\begin{equation}
	\mathcal{M} = \left\{ \zeta:  \zeta = \left(\vec{H},\vec{B}\right) \in \mathrm{H}(\text{curl}) \times \mathrm{H}(\text{div}):\eqref{eq:magnetostatic} , \mathrm{a.e.} \in \Omega \right\}.
	\label{eq:maxwell_set}
\end{equation}
We further define the global material data set 
\begin{equation}
	\mathcal{D} = \left\{\zeta :\zeta \in  \mathrm{L}^2(\Omega)^3\times \mathrm{L}^2(\Omega)^3, \,\zeta(\vec{x}) \in \widetilde{\mathcal{D}},\,\mathrm{a.e.} \in \Omega\right\}.
	\label{eq:material_set}
\end{equation}
The solution of the data-driven solver is then obtained as
\begin{equation}
	\mathcal{S} = \argmin_{\zeta \in \mathcal{M}} \left\{ F(\zeta,\mathcal{D})\right\},
	\label{eq:data_driven_sol}
\end{equation}
where $F(\zeta,\mathcal{D}) = \mathrm{inf} \{||\zeta-\zeta^\star||:\zeta^\star \in \mathcal{D}\}$ is a distance function in the $HB$-phase space.
Let $(\vec{u},\,\vec{v})_\Omega = \int_\Omega \vec{u} \cdot \vec{v}\, \mathrm{d}\Omega$, where $\vec{u},\, \vec{v} \in \mathrm{L}^2(\Omega)^3$ denote the inner product. Then, a possible choice for the distance function is given by
\begin{equation}
	F(\zeta,\mathcal{D}) 
	= \min_{\zeta^\star \in \mathcal{D}} f(\zeta,\zeta^\star) = \min_{\zeta^\star \in \mathcal{D}}\frac{1}{2} \left(\vec{H} - \vec{H}^\star,\widetilde{\boldsymbol{\mu}}  (\vec{H} - \vec{H}^\star)\right)_\Omega + \frac{1}{2} \left(\vec{B} - \vec{B}^\star,  \widetilde{\boldsymbol{\nu}}  (\vec{B} - \vec{B}^\star)\right)_\Omega,
	\label{eq:dist_func}
\end{equation}
where $\widetilde{\boldsymbol{\mu}}$, respectively $\widetilde{\boldsymbol{\nu}}$, are weighting factors in the $HB$-phase space. 
Essentially, \eqref{eq:data_driven_sol} is a double minimization problem, i.e.
\begin{equation}
	\mathcal{S} = \argmin_{\zeta \in \mathcal{M}} \left\{ \min_{\zeta^\star \in \mathcal{D}} \left\{ f(\zeta,\zeta^\star) \right\}\right\},
\end{equation}
which tries to match Maxwell-conforming states to states described by the given material data.
In the rest of this work, we consider only the case of a diagonal material tensor. To ensure a convex minimization problem, the weighting factors are bounded by
\begin{subequations}
	\begin{alignat}{3}
		&\mu_0 &&\le \widetilde{\mu}_d(\vec{x}) < \infty, \\
		&0 &&< \widetilde{{\nu}}_d(\vec{x}) \le \nu_0,
	\end{alignat}%
	\label{eq:mu_constraints}%
\end{subequations}%
where $\mu_0$ and $\nu_0$ refer to the permeability, respectively reluctivity in vacuum and $d\in\{x,y,z\}$. Furthermore, the weighting factors are of computational nature, i.e. they do not have to fulfill any physical constraints except \eqref{eq:mu_constraints}. 

To solve the minimization problem, we express the magnetic field strength as $\vec{B} = \mathrm{curl}\,\vec{A}$. Amp\`{e}re's law is enforced through the Lagrange multiplier $\vec{\eta}$, which leads to the Lagrangian
\begin{equation}
	\mathcal{L}\left(\vec{H},\vec{A},\vec{\eta}\right)=F\left((\vec{H},\mathrm{curl}\, \vec{A}),\zeta^\star\right) - \left(\vec{\eta}, \vec{J} - \mathrm{curl}\, \vec{H}\right)_\Omega.
	\label{eq:lagrangian}
\end{equation}
\rev{The Lagrangian \eqref{eq:lagrangian} is solved in a two-step minimization scheme. First, \eqref{eq:lagrangian} is minimized over $\vec{A},\vec{H},\vec{\eta}$ for a specific choice of $\zeta^\star \in \mathcal{D}$, denoted with the superscript $^\times$. The obtained state $\zeta$ is then compatible with Maxwell's equations, i.e. $\zeta \in \mathcal{M}$. Afterwards, the recently obtained state $\zeta \in \mathcal{M}$ is minimized towards the available measurement data by solving \eqref{eq:dist_func}. The result, denoted by $\zeta^\times \in \mathcal{D}$ is thus closest to the Maxwell conforming state $\zeta$ and again employed in \eqref{eq:lagrangian}. This two-step minimization scheme is then iterated until $\zeta^\times$ does not change after two consecutive iterations.}

As we seek for the stationary points of \eqref{eq:lagrangian}, we take the \emph{functional derivatives} \cite{gelfand1963} $\nabla_{\kern -0.2em A}, \nabla_{\kern -0.2em H}, \nabla_{\kern -0.2em \eta}$, such that 
\begin{subequations}
	\begin{alignat}{4}
		&\nabla_{\kern -0.2em A} &\mathcal{L} &: \quad &\mathrm{curl}\,\left(\widetilde{\boldsymbol{\nu}}\mathrm{curl}\,\vec{A}\right) - \mathrm{curl}\,\left(\widetilde{\boldsymbol{\nu}}\vec{B}^\times\right) &= 0, \label{eq:dA}\\ 
		&\nabla_{\kern -0.2em H}  &\mathcal{L}&:   \quad  &\widetilde{\boldsymbol{\mu}}\left(\vec{H} - \vec{H}^\star\right)+\mathrm{curl}\,\vec{\eta} &= 0, \label{eq:dH}\\
		&\nabla_{\kern -0.2em \eta} &\mathcal{L}&:\quad  &\vec{J} - \mathrm{curl}\,\vec{H} &= 0. \label{eq:deta}
	\end{alignat} %
	\label{eq:lagragne_derivative}%
\end{subequations}
\rev{The \gls{fem} is utilized to solve \eqref{eq:lagragne_derivative}. Therefore, the stationary solution \eqref{eq:lagragne_derivative} must be cast  into the weak formulation. Applying the curl operator on \eqref{eq:dH} and inserting \eqref{eq:deta}, we obtain a curl-curl equation on the Lagrangian $\vec{\eta}$. Now, the recently obtained curl-curl equation acting on $\vec{\eta}$ as well as the curl-curl equation \eqref{eq:dA} acting on $\vec{A}$ can be cast into the weak formulation.}
\rev{Finally}, we seek for $\vec{A},\vec{\eta}\in V=\{\vec{v} \in \mathrm{H}(\text{curl}): \vec{v}\times \vec{n}=0 \text{~on~}\Gamma\}$, such that
\begin{subequations}
	\begin{align} 
		a(\vec{A},\vec{w})    &= l_1^\times(\vec{w}),  \quad \forall \vec{w}\in V, \\
		a(\vec{\eta},\vec{w}) &= l_2^\times(\vec{w}),  \quad \forall \vec{w}\in V,
	\end{align}
	\label{eq:weak_form_with_zero_kernel}%
\end{subequations}
where the bilinear forms are given by
\begin{subequations}
	\begin{align}
		a(\vec{A},\vec{w}) &:= (\widetilde{\boldsymbol{\nu}}\mathrm{curl}\vec{A},\mathrm{curl}\vec{w})_\Omega, \\
		a(\vec{\eta},\vec{w}) &:= (\widetilde{\boldsymbol{\nu}}\mathrm{curl}\vec{\eta},\mathrm{curl}\vec{w})_\Omega,
	\end{align}
	\label{eq:bilinear_form}%
\end{subequations}
and the \glspl{rhs} by
\begin{subequations}
	\begin{align}
		l_1^\times(\vec{w})&=(\widetilde{\boldsymbol{\nu}} \vec{B}^\times,\mathrm{curl}\vec{w})_\Omega, \label{eq:rhs1} \\
		l_2^\times(\vec{w})&= (\vec{H}^\times,\mathrm{curl}\vec{w})_\Omega - (\vec{J},\vec{w})_\Omega. \label{eq:rhs2}
	\end{align}
	\label{eq:rhss}%
\end{subequations}
Note that the superscript $^\times$ indicates that a particular state $\zeta^\times \in \mathcal{D}$ has been chosen.

In the presented form, the weak formulation \eqref{eq:weak_form_with_zero_kernel} features a large null space. As a remedy, we employ the Coulomb gauge $\mathrm{div}\,\vec{A}=0$, for which we dispense a detailed exposition and refer the reader to the relevant literature, e.g. \cite{biro2007CoulombGauge}.
After solving \eqref{eq:weak_form_with_zero_kernel}, a new field solution for $\vec{H}$ is computed with \eqref{eq:dH}, whereas $\vec{B}$ is obtained by post-processing $\vec{A}$. The updated terms read
\begin{subequations}
\begin{align}
	\vec{B} &= \mathrm{curl}\vec{A}, \\
	\vec{H} &= \vec{H}^\times + \widetilde{\boldsymbol{\nu}} \mathrm{curl}\vec{\eta}.		
\end{align}
\label{eq:update_field}%
\end{subequations}
With \eqref{eq:update_field} we have obtained a state $\zeta \in \mathcal{M}$. The next step in the data-driven algorithm is to find a state $\zeta^\times \in \mathcal{D}$, i.e. a state in the measurement set, that is closest to our recently updated Maxwell-conforming state $\zeta \in \mathcal{M}$, which is then employed in the \glspl{rhs} \eqref{eq:rhss} for the next iteration step. This two-step minimization procedure, i.e. first minimizing in the Maxwell-conforming set $\mathcal{M}$ and afterwards in the measurement-conforming set $\mathcal{D}$, is continued until convergence is reached. 

Next, we take a closer look at the second minimization step. Constructing all possible states $\zeta^\times \in \mathcal{D}$ to perform the minimization would be computationally unfeasible \cite{nguyen2020_ddframework}. Instead, a quadrature scheme is employed to compute the integral in \eqref{eq:dist_func}.
Let $\Delta\vec{H} = \vec{H}-\Hm$ and $\Delta\vec{B}=\vec{B}-\Bm$ denote the distances between two states. We seek the state $\zeta^\times \in \mathcal{D}$ that minimizes \eqref{eq:dist_func} at the quadrature points, i.e.
\begin{equation}
	\boldsymbol{\zeta}^\times(\vec{x}_{\qd}) =  \argmin_{\zeta^\star \in \mathcal{D}}\frac{1}{2}\sum_{i=1}^{N_\mathrm{qd}}w_i \Delta\vec{H}(\vec{x}_{\qd,i}) \cdot \tilde{\boldsymbol{\mu}}(\vec{x}_{\qd,i})\Delta\vec{H}(\vec{x}_{\qd,i}) \\ + \Delta\vec{B}(\vec{x}_{\qd,i}) \cdot \tilde{\boldsymbol{\nu}}(\vec{x}_{\qd,i})\Delta\vec{B}(\vec{x}_{\qd,i}),
	\label{eq:min_discrete}
\end{equation}
where $N_\qd$ denotes the number of quadrature points, $\vec{x}_{\qd,i}$ a quadrature point, $w_i$ a quadrature weight and $\boldsymbol{\zeta}^\times(\vec{x}_\qd) = (\zeta^\times(\vec{x}_{\qd,i}),\dots,\zeta^\times(\vec{x}_{\qd,N_\qd}))$. Note that, in the discretized case, the evaluation of \eqref{eq:dist_func} with \eqref{eq:min_discrete} is exact for a sufficiently high quadrature degree. 

\textbf{Remark 1} The considered example features domains where the material behavior is only given by measurement data, and domains where the constitutive relation is known, i.e. in the case of vacuum. In the case of a domain with a known constitutive law, the minimization \eqref{eq:min_discrete} is performed with the known closed-form relation of the material law, as described in \cite{degersem2020magnetic}. Accordingly, the weighting factors $\tilde{\boldsymbol{\mu}}$, respectively $\tilde{\boldsymbol{\nu}}$, are equal to the known material coefficients.

\textbf{Remark 2} A recent work by the authors \cite{galetzka2020datadriven} showed that heterogeneous (local) weighting factors $\tilde{\boldsymbol{\mu}}$, respectively $\tilde{\boldsymbol{\nu}}$, improve significantly the accuracy and efficiency of the data-driven solver in the case of unbalanced material data sets. A local weighting factor is adaptively matched to the current operation point of the field state at the quadrature point. More precisely, the weighting factor is chosen to be the local tangent of the current state, with respect to the surrounding states in the material data set $\mathcal{D}$. With the local weighting factors, the data-driven solver compensates for information loss due to data-starved or unevenly filled material data sets.

\section{Numerical results}
In the following, a number of numerical experiments are presented, aiming to showcase the performance and accuracy of the data-driven solver. First, we validate that the data-driven solver recovers the solution obtained with a conventional Newton solver when a data set of increasing cardinality is employed. Second, a data set containing real measurements is utilized, showing the applicability of the data-driven solver in this practical test case.

\begin{figure}[t!]
	\begin{minipage}[t]{0.39\textwidth}
		\centering
		\includegraphics[width=0.7\textwidth]{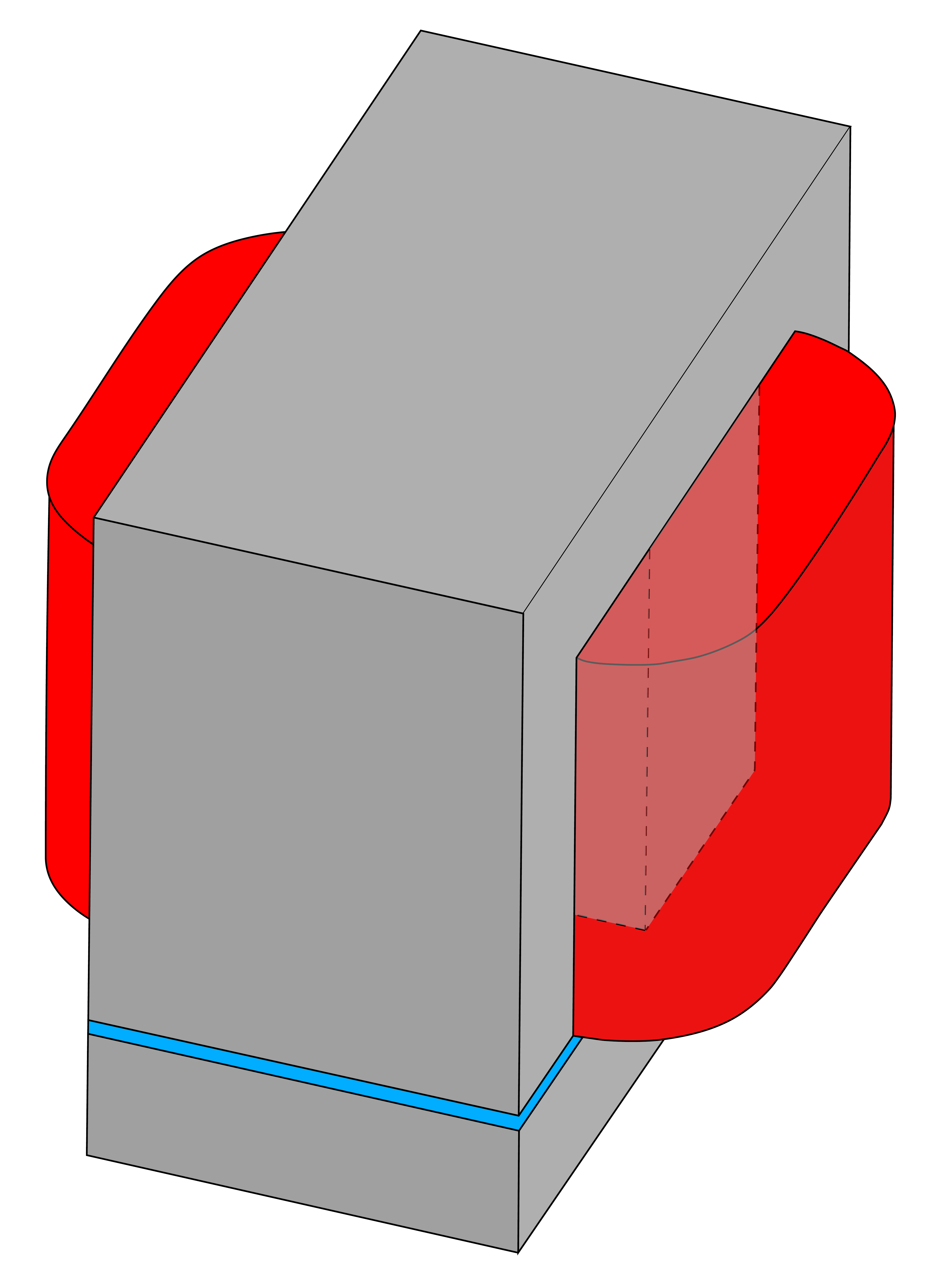} %0.6
		\caption{Full model of the inductor with iron- (gray), coil- (red) and air-region (blue).}
		\label{fig:inductor_model}
	\end{minipage}
	\hfill
	\begin{minipage}[t]{0.6\textwidth}
		\centering
		\includegraphics[width=1.00\textwidth]{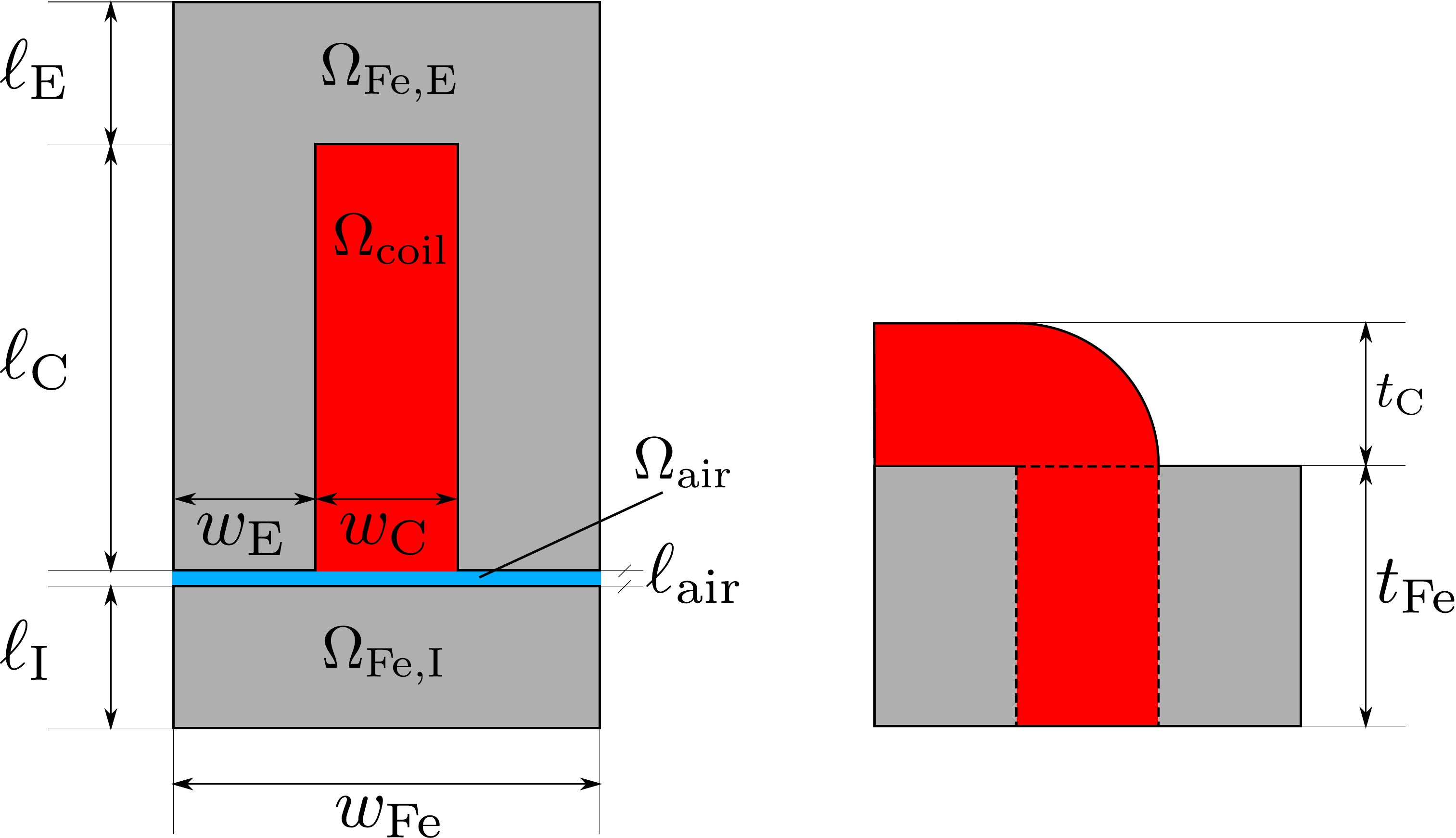} %0.95
		\caption{$2$D cross-section of the inductor model.}
		\label{fig:inductor_model_2D}
	\end{minipage}		
\end{figure} 

\begin{table}[b!]
\centering
\begin{minipage}[t]{0.48\linewidth}\centering		
\caption{Geometrical dimensions of the inductor model.} 
%	\ra{1.3}
\centering
\raisebox{\depth}{\begin{tabular}[htbp]{cc}
		\toprule
		quantity & value in mm  \\
		\midrule
		$\ell_\mathrm{E}$      & $30$  \\
		$\ell_\mathrm{C}$      & $90$ \\ 
		$\ell_\mathrm{I}$      & $30$   \\ 
		$\ell_\mathrm{air}$    & $3.3$   \\ 
		$w_\mathrm{E}$         & $30$   \\ 
		$w_\mathrm{C}$         & $30$   \\ 
		$w_\mathrm{Fe}$         & $90$   \\ 
		$t_\mathrm{C}$         & $30$   \\ 
		$t_\mathrm{Fe}$         & $55$   \\ 
		\bottomrule
\end{tabular}}
\label{tab:inductor_geom}
\end{minipage}\hfill%
\begin{minipage}[t]{0.48\linewidth}\centering
	\caption{Material properties and excitation.} 
	%	\ra{1.3}
	\centering
	\raisebox{\depth}{\begin{tabular}[htbp]{ccc}
			\toprule
			quantity & value & units \\
			\midrule
			$k_1$      & $10$  & $\SI{}{\meter\per\henry}$\\
			$k_2$      & $1.8$ &  $\SI{}{\per\square\tesla}$ \\
			$k_3$      & $100$  & $\SI{}{\meter\per\henry}$\\
			$I$        & $50$ & $\SI{}{\ampere}$\\
			$N_\mathrm{coil}$ & $66$ & --\\
			\bottomrule
	\end{tabular}}
	\label{tab:inductor_param}
\end{minipage}
\end{table}

In all numerical experiments, we consider the three-dimensional model of a \rev{DC-current electromagnet}, see Figure~\ref{fig:inductor_model}. The inductor consists of an E-shaped iron yoke part $\Omega_{\mathrm{Fe,E}}$, inside which a wire winding is arranged. The coil region is denoted by $\Omega_\mathrm{coil}$. The magnetic circuit is closed by an I-shaped yoke part $\Omega_{\mathrm{Fe,I}}$, however, only up to an air gap of length $\ell_\mathrm{air}$, which defines the domain $\Omega_{\mathrm{air}}$. Details about the geometrical dimensions can be found in Table~\ref{tab:inductor_geom} and in Figure~\ref{fig:inductor_model_2D}.
Furthermore, the inductor is surrounded by an air-filled sphere, the domain of which is denoted with $\Omega_{\mathrm{sphere}}$. The radius of the sphere is given by $r_\mathrm{s} = \SI{0.28}{\meter}$.

Since the magnetostatic problem is formulated through the vector potential, we apply magnetic wall \glspl{bc}, i.e. homogeneous Dirichlet \glspl{bc}, on the sphere boundary. Because the radius of the sphere is an order of magnitude larger than the inductor model, no notable approximation error is introduced. Exploiting the symmetry of the inductor, it is sufficient to simulate only one fourth of its geometry, reducing the computational costs drastically. Then, homogeneous Dirichlet \glspl{bc} are applied on the boundaries related to the symmetry. The excitation current of the coil is modeled by ``stranded conductors'', \cite{bedrosian1993}. By that, we assume that the current density is constant across the domain of the coil, such that the computationally expensive modeling of the single wires of the coil is circumvented.

For the discrete weak formulation, first order curl-conforming N\'{e}d\'{e}lec functions \cite{nedelec1980} are employed as ansatz and test functions for $\vec{A},\vec{\eta}$, and $\vec{w}$. The domain is discretized into $193311$ elements, leading to $263627$ \glspl{dof}. The corresponding numbers of edges and vertices are $230375$ and $33252$, respectively. \rev{The reference solutions are calculated with the Python library FEniCS \cite{Blechta2015a_fenics}. The data-driven solver is a modification thereof.}

\subsection{Performance and accuracy}
\label{sec:numerical_results}

A nonlinear, isotropic material behavior $H\left(B\right)$ is considered for the two yoke parts, following the Brauer model \cite{brauer1975}. For each dimension, the Brauer model is defined by
\begin{equation}
	H(B) = \nu(B)B = \left(k_1 \mathrm{e}^{k_2B^2}+k_3\right) B.
\end{equation}
The constants defining the material law are summarized in Table~\ref{tab:inductor_param}, along with additional parameters which define the inductor model.
Yet, the data-driven solver employs only a finite number of sampled $BH$-pairs of the model. The material model is sampled equidistantly for each component of $B$. However, due to the nonlinearity in the material model, the equidistance property is lost for the corresponding magnetic field strength data. 

Assuming an infinite number of measurement points, the data-driven solver should converge to the classical solution, e.g. obtained with a Newton solver \cite{kirchdoerfer2016data}. 
For validation purposes, we simulate the inductor model for data sets $\mathcal{D}$ of increasing cardinality. The error of the data-driven solver towards the reference solution is measured with the energy mismatch 
\begin{equation}
	\epsilon_{\mathrm{em}}^2 = \frac{\left(\vec{H}-\vec{H}_\mathrm{ref},\boldsymbol{\mu}_\mathrm{ref}(\vec{H}-\vec{H}_\mathrm{ref})\right)_\Omega }{\left(\vec{H}_\mathrm{ref},\boldsymbol{\mu}_\mathrm{ref}\vec{H}_\mathrm{ref}\right)_\Omega + \left(\vec{B}_\mathrm{ref},\boldsymbol{\nu}_\mathrm{ref}\vec{B}_\mathrm{ref}\right)_\Omega} +\frac{\left(\vec{B}-\vec{B}_\mathrm{ref},\boldsymbol{\nu}_\mathrm{ref}(\vec{B}-\vec{B}_\mathrm{ref})\right)_\Omega}{\left(\vec{H}_\mathrm{ref},\boldsymbol{\mu}_\mathrm{ref}\vec{H}_\mathrm{ref}\right)_\Omega + \left(\vec{B}_\mathrm{ref},\boldsymbol{\nu}_\mathrm{ref}\vec{B}_\mathrm{ref}\right)_\Omega},
\end{equation}
where $\boldsymbol{\mu}_\mathrm{ref}$ and $\boldsymbol{\nu}_\mathrm{ref}$ refer to the permeability and reluctivity obtained with the reference solution.

\begin{figure}[t!]
	\begin{minipage}[t]{0.49\textwidth}
		\centering
		\includegraphics[width=0.8\textwidth]{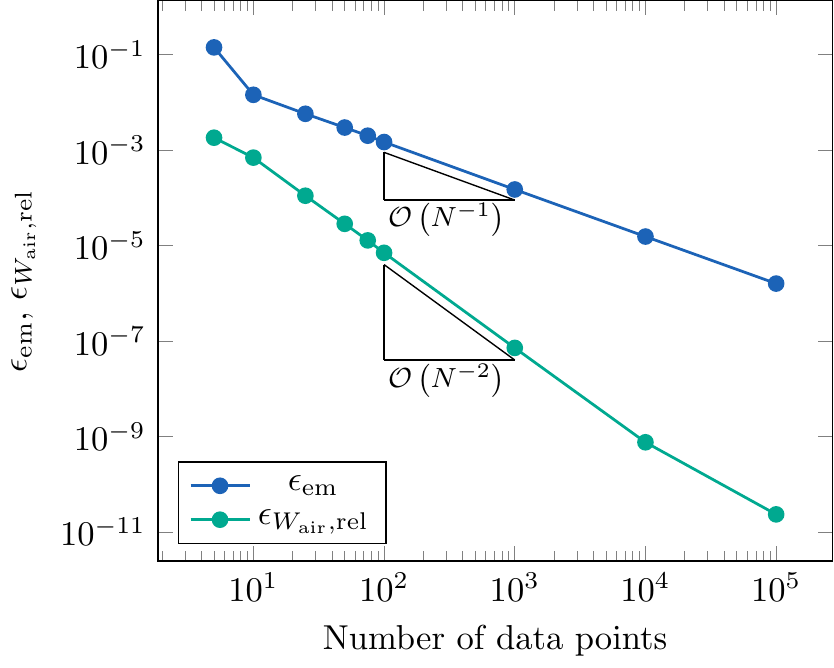}
		\caption{Convergence of the energy mismatch for increasing data set size.}
		\label{fig:rms_distance_to_ref}
	\end{minipage}
	\hfill
	\begin{minipage}[t]{0.49\textwidth}
		\centering
		\includegraphics[width=0.8\textwidth]{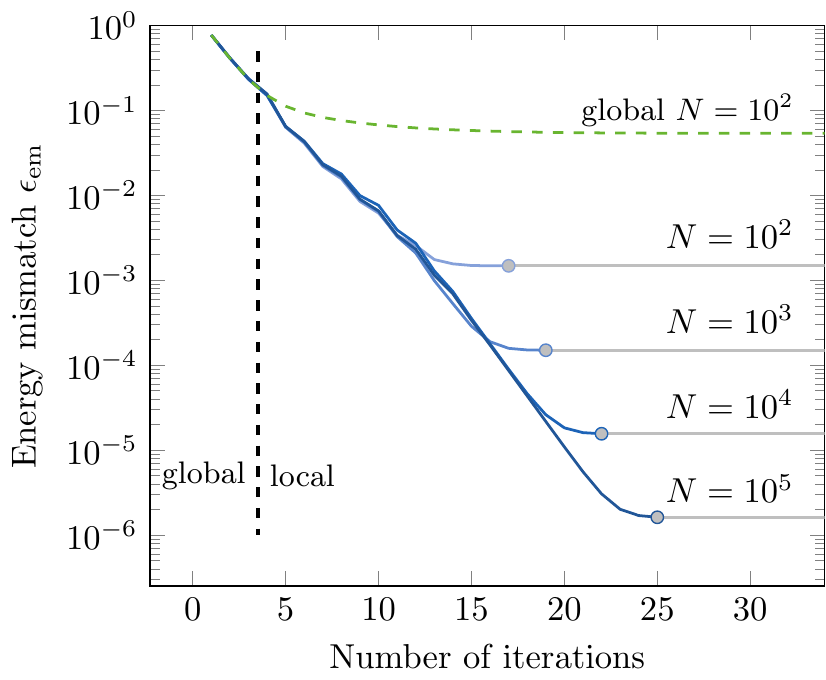}
		\caption{Convergence of the energy mismatch in dependence of the number of data-driven iterations. The green dashed line shows the results without switching to a local weighting factor.}
		\label{fig:rms_dist_over_iterations}
	\end{minipage}	
\end{figure} 

Figure~\ref{fig:rms_distance_to_ref} shows the energy mismatch $\epsilon_{\mathrm{em}}$ in dependency to the cardinality of $\mathcal{D}$. We can clearly observe that the data-driven solver converges towards the reference solution as the number of measurement points increases. Furthermore, we achieve linear convergence with respect the number of data points. The convergence with respect to the number of data-driven iterations is shown in Figure~\ref{fig:rms_dist_over_iterations}. The switching point from a global (homogeneous) weighting factor to local (heterogeneous) weighting factors was fixed to $4$ iterations. Selecting the switching point adaptively is also possible \cite{galetzka2020datadriven}.

Here, the data-driven solver's convergence stagnates due to the finite number of employed measurement data points. We also observe that the cardinality of $\mathcal{D}$ affects the number of solver iterations until convergence is reached. Furthermore, the convergence of the energy mismatch is tremendously improved when switching from global to local weighting factors. This can be attributed to the unbalanced measurement set $\mathcal{D}$, i.e. to the (locally) sparse data. The data-driven solver with a global weighting factor faces major difficulties to produce accurate solutions, since it operates only on one operation point on the $BH$-curve \cite{galetzka2020datadriven}.

\begin{figure}[t!]
	\begin{minipage}[t]{0.49\textwidth}
		\centering
		\includegraphics[width=0.85\textwidth]{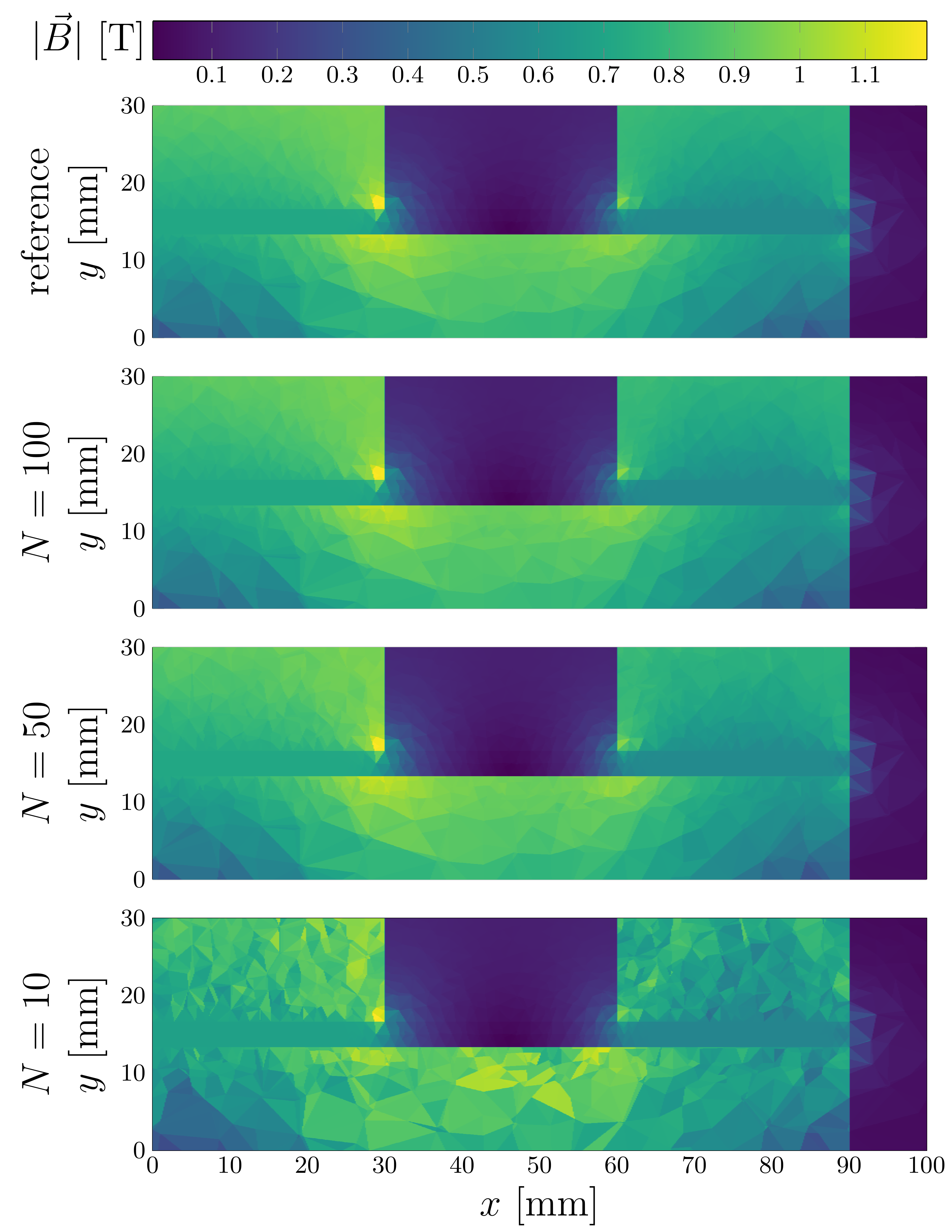}
		\caption{\rev{Magnitude of the magnetic flux density at the air gap for the reference solution (top) and the data-driven solutions computed with $N\in \{10,50,100\}$.}}
		\label{fig:Bfield_airgap}
	\end{minipage}
	\hfill
	\begin{minipage}[t]{0.49\textwidth}
		\centering
		\includegraphics[width=1.025\textwidth]{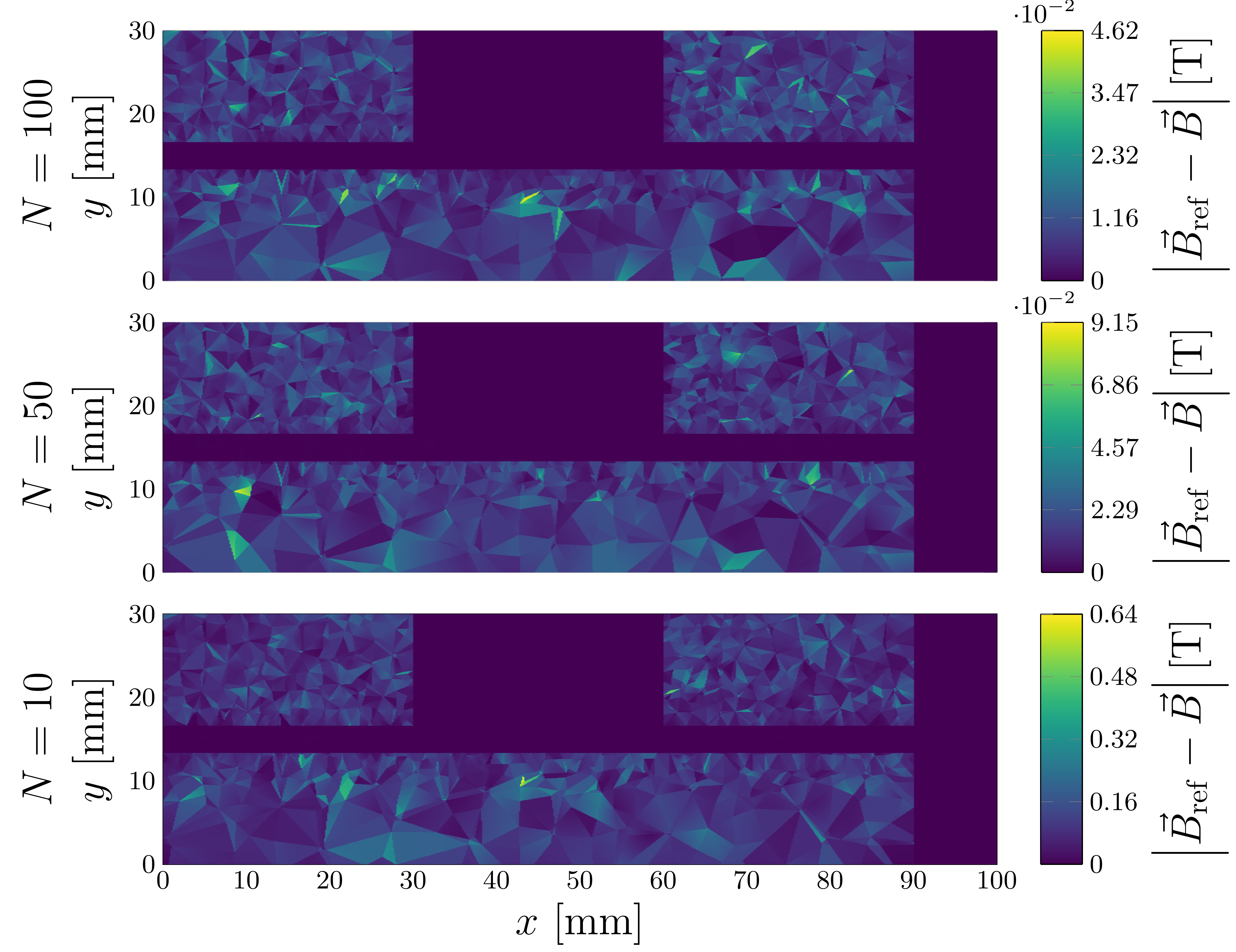}
		\caption{\rev{Absolute difference in the magnetic flux density between the reference solution and the data-driven solutions computed with $N\in \{10,50,100\}$.}}
		\label{fig:Bfield_diff}
	\end{minipage}	
\end{figure}

As a local measure of accuracy, we compute the magnetic energy in the air gap, which is given by
\begin{equation}
	W_\mathrm{air} = \int_{\Omega_{\mathrm{air}}} \frac{1}{2} \vec{H} \cdot \vec{B} \,\mathrm{d}\Omega_{\mathrm{air}}.
\end{equation}
The corresponding relative error in the energy is
\begin{equation}
\epsilon_{W_\mathrm{air},\mathrm{rel}} = \frac{\left|W_\mathrm{air} - W_\mathrm{air,ref} \right|}{W_\mathrm{air,ref}},
\end{equation}
where the $W_\mathrm{air,ref}$ is calculated from the reference solution. The convergence of the energy in the air gap is also shown in Figure~\ref{fig:rms_distance_to_ref} and exhibits a quadratic convergence with respect to the cardinality of the measurement set. Furthermore, Figure~\ref{fig:Bfield_airgap} shows the magnitude of the magnetic flux \rev{density} for the reference solution as well as for data-driven solutions computed with \rev{$N\in\{10,50,100\}$} measurement samples. The absolute difference, i.e. $|\vec{B}_\mathrm{ref}-\vec{B}|$ is shown in Figure~\ref{fig:Bfield_diff}. Both the local error shown in Figure~\ref{fig:Bfield_diff} and the relative error in the energy of the air gap, show that for a measurement set of $N=10^2$ data points, a sufficient accuracy with $\epsilon_{W_{\mathrm{air}},\mathrm{rel}}\approx7 \cdot10^{-6}$ is obtained. \rev{Figures~\ref{fig:Bfield_airgap} and \ref{fig:Bfield_diff} also show that if only $N=10$ data points are employed, the field solution is not adequately resolved and the absolute field error is beyond standard engineering tolerances.}

From the results it is concluded that the data-driven magnetostatic solver can be employed to solve real-world three-dimensional electromagnetic problems. The data-driven solver's computational cost is dominated by the solutions of the linear systems, as is the case for a standard Newton solver. However, the data-driven solver must solve two linear systems in each iteration instead of just one. Moreover, if large data sets are available, e.g. with more than $N=10^5$ data points, the computational cost of minimizing \eqref{eq:min_discrete} in a brute-force manner is not negligible anymore. A possible remedy would be to perform a localized minimization, i.e. for a given state $\zeta(\vec{x}_{\qd,i}) \in \mathcal{M}$ only surrounding states in $\mathcal{D}$ are employed to solve \eqref{eq:min_discrete}, instead of the entire data set.
Taking into account that the optimal computational complexity when solving the linear systems by a multigrid solver scales linearly with the number of \glspl{dof} \cite{trottenberg2001multigrid}, we conclude that the data-driven solver scales as favorably as a standard Newton solver.

For the considered example, a conventional Newton solver reaches machine accuracy with approximately $10$ iterations. Contrarily, the data-driven solver needs roughly $20$ iterations, each invoking two system solves, to reach an accuracy of $10^{-6}$. Still, this is affordable, especially in cases where one exploits the advantage of a data-driven solver, which is the cancellation of epistemic errors related to a possibly non-adequate choice of material model.

\subsection{Real world measurement data}
\label{sec:numerical_results_real_world}
\begin{figure}[t!]
	\begin{minipage}[t]{0.49\textwidth}
		\centering
		\includegraphics[width=1.0\textwidth]{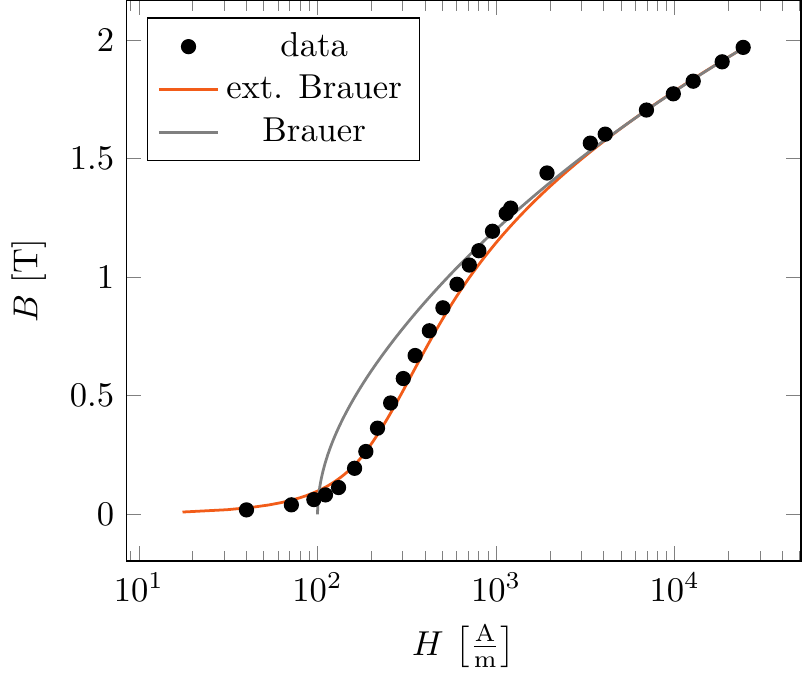}
		\caption{The measurement data of the S355 sample is shown in black, the standard Brauer model in gray, and the extended Brauer model in orange.}
		\label{fig:S355_steel}
	\end{minipage}
	\hfill
	\begin{minipage}[t]{0.49\textwidth}
		\centering
		\includegraphics[width=1.0\textwidth]{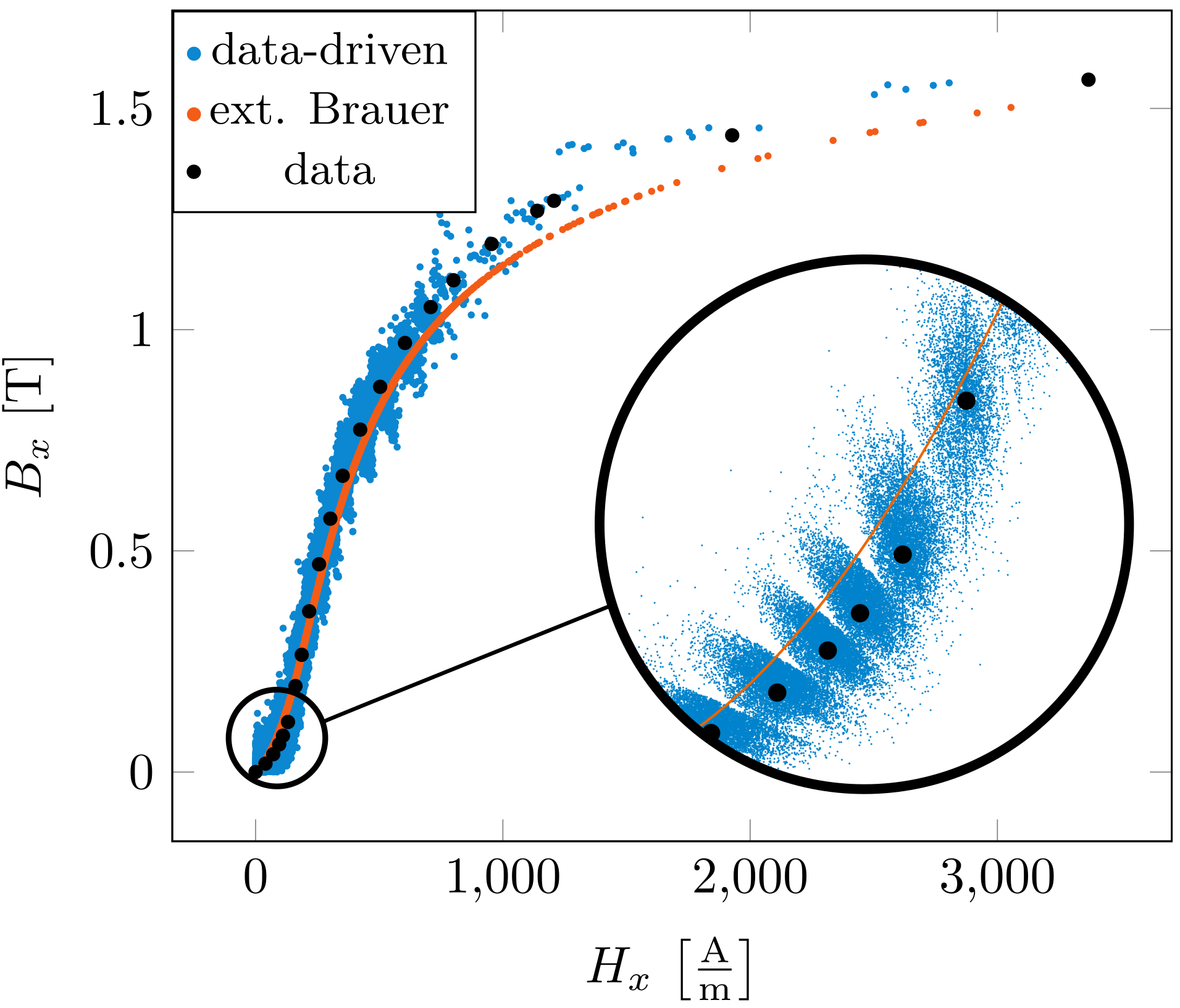}
		\caption{Solution of the data-driven solver (blue) and of the conventional solver employing the extended Brauer (orange) at the quadrature points. The measurement data is shown in black.}
		\label{fig:res_S355}
	\end{minipage}	
\end{figure}

In the previous section we have shown that the data-driven solver recovers the conventional solution. Yet, its strength lies in cases where little or no information about the material law is available, apart from measurement data. 
Therefore, for the next simulations, the data set $\mathcal{D}$ contains only a fixed number of real-world measurement data. 
In particular, we consider a low-cost
structural steel, called S355. 
The employed measurement data is taken from \cite{anglada2020}. 
In addition to the data-driven solution, a solution obtained with a conventional solver is computed as well. For the latter, a material law fitted to the available measurement data needs to be constructed. We consider the extended Brauer model \cite{huelsmann2014}, which improves the standard Brauer model in the Rayleigh region, as well as in the region of saturation. The extended Brauer model together with the measurement data is depicted in Figure~\ref{fig:S355_steel}. Additionally, the Brauer model is shown, exhibiting good approximation characteristics in the saturation part, yet the model is not capable to resolve the Rayleigh region. However, both methods show approximation errors and, due to the nature of the model assumptions, feature additional epistemic uncertainties.

Since the data can be considered as noise-free, we abandon the introduction of the noisy data-driven solver \cite{kirchdoerfer2017data, galetzka2020datadriven} and employ the standard, distance-minimizing data-driven solver introduced in Section~\ref{sec:dd_framework}. The excitation current is adjusted to the bandwidth of the available measurement data and set to $I=\SI{40}{\ampere}$. 

As no reference solution is available, we analyze the results of both methods qualitatively first. Figure~\ref{fig:res_S355} shows exemplarily the magnitude of the magnetic field strength and magnetic flux \rev{density} in the $x$-direction, i.e. $(|H_x|,|B_x|)$, at the quadrature points of the \gls{fe}-procedure. The figure depicts both the data-driven solution as well as the conventional solution calculated with the extended Brauer model. We observe that the data-driven solution builds clusters around the measurement data, which is to be expected due to the sparse data set consisting of only $28$ measurements. Nevertheless, the data-driven solution remains in the area around the given measurement data points in \emph{all} regions of the $BH$-curve. Contrarily, the conventional solution shows a good agreement with the measurement data in the steep part of the $BH$-curve, but is comparatively inaccurate in the saturation and the Rayleigh parts. Still, both methods suffer from the sparse data set, albeit differently.

To quantify the accuracy of the solutions, we calculate the energy mismatch $\epsilon_{\mathrm{em,data}}$ between the field solutions and the closest measurement states, i.e. states $(\vec{H}^\times,\vec{B}^\times) \in \mathcal{D}$ that minimize \eqref{eq:min_discrete} for the given solution. The energy mismatch then reads
\begin{equation}
	\epsilon_{\mathrm{em,data}}^2 = \left(\vec{H}-\vec{H}^\times,\boldsymbol{\mu}_\mathrm{ext.Brauer}(\vec{H}-\vec{H}^\times)\right)_\Omega +\left(\vec{B}-\vec{B}^\times,\boldsymbol{\nu}_\mathrm{ext.Brauer}(\vec{B}-\vec{B}^\times)\right)_\Omega,
\end{equation}
where we employed the approximated permeability, respectively reluctivity of the extended Brauer model. The energy mismatch for the data-driven solution is given by $\epsilon_{\mathrm{em,dd}}=2.9\cdot 10^{-4}\si{\joule}$, whereas the corresponding error for the conventional solution is $\epsilon_{\mathrm{em,dd}}=3.9\cdot 10^{-4}\si{\joule}$. The two approaches show comparable accuracy.

In summary, both approaches lead to an acceptable solution for the given measurement data. However, the material characteristic of the sample is well studied, that is, significant effort has already been spent in the model assumptions. The conventional solver benefits from this modeling information, while the data-driven solver disregards it altogether and yields an assumptions-free solution.
It can therefore be concluded that, considering more complex materials with little to no modeling information available, the data-driven solver should be the superior solution method.

\section{Conclusion}
\label{sec:conclusion}
A magnetostatic data-driven solver was presented and successfully applied to a real-world three-dimensional test case. The solver recovers the solution obtained with a conventional Newton solver when more measurement data are utilized.
The data-driven solver constitutes a viable alternative to conventional Newton solvers when no explicit material model exists and only material measurement data are available instead. 
Even if material models are available, the utilization of a data-driven solver avoids so-called epistemic uncertainties, as only raw measurement data are used for the solution.

\section*{Acknowledgment}
This work has been supported by the DFG, Research Training Group 2128 ''Accelerator Science and Technology for Energy Recovery Linacs''. 
The work of D. Loukrezis is further supported by the BMBF via the research contract 05K19RDB.

\bibliographystyle{agsmdoi} 
\bibliography{references}

\end{document}